\pdfoutput=1
\documentclass[sigplan,screen,10pt,nonacm]{styles/acmart-primary/acmart}

\usepackage{xspace}
\usepackage{algorithm}
\usepackage[noend]{algorithmic}

\newcommand{\projectname}{our system\xspace}
\newcommand{\Projectname}{Our system\xspace}   
\newcommand{\predictor}{Predictor\xspace}
\newcommand{\constructor}{Constructor\xspace}
\newcommand{\scheduler}{Scheduler\xspace}
\newcommand{\projtitle}{Latency-Aware Orchestration for Multi-Agent LLM Workflows on Heterogeneous GPUs}

\title[]{\projtitle}

\setcopyright{none}
\author{%
  \mbox{Jinghao Wang\textsuperscript{1}},
  \mbox{Yifeng Zhang\textsuperscript{1}},
  \mbox{Xiao Zhou\textsuperscript{1}},
  \mbox{Yao Lu\textsuperscript{1}},
  \mbox{Yihui Zhang\textsuperscript{1}},
  \mbox{Xiaoyang Sun\textsuperscript{2}},
  \mbox{Tianyu Wo\textsuperscript{1}},
  \mbox{Xu Wang\textsuperscript{1}},
  \mbox{Chunming Hu\textsuperscript{1}},
  \mbox{Renyu Yang\textsuperscript{1}}}
\affiliation{%
  \institution{\textsuperscript{1}Beihang University, Beijing, China \qquad \textsuperscript{2}University of Leeds, Leeds, United Kingdom}
  \city{}
  \country{}}

\renewcommand{\shortauthors}{Wang et al.}

\begin{document}

\begin{abstract}
Concurrent multi-agent workflows expose future dependencies and serving-state requirements while running on heterogeneous GPU pools with time-varying load, model residency, and resource availability. The logical workflow defines the required computation, whereas its physical scheduling units, model-lifecycle actions, resource ordering, and placement must be selected according to the observed pool state.
We present a prediction-guided runtime that uses workflow forecasts to construct and optimize a physical execution graph.
\predictor{} estimates device-specific activation latency, peak memory, and model-loading cost, then propagates these predictions through workflow dependencies to forecast activation readiness and future model demand.
\constructor{} builds semantics-preserving fusion and model-lifecycle alternatives, while \scheduler{} jointly optimizes their selection, placement, and execution order based on the live pool state.
Across a workload spanning three workflow scenarios on a heterogeneous GPU pool, \projectname{} reduces end-to-end makespan and overall p95 completion latency under burst arrivals by up to 36.8\% and 25.9\%, respectively, over state-of-the-art workflow schedulers. It also saves up to 24.63 GPU-s per completed session.

\end{abstract}


\maketitle

\section{Introduction}
\label{sec:introduction}

LLMs such as DeepSeek~\cite{deepseek2025r1}, Qwen~\cite{yang2025qwen3}, and LLaMA~\cite{touvron2023llama} increasingly power compound AI systems that orchestrate model calls, tools, and control flow~\cite{zaharia2024compound}.
LangGraph~\cite{langchain2026langgraph}, AutoGen~\cite{wu2023autogen}, and AgentScope~\cite{gao2024agentscope} express these applications as stateful Agent graphs.
Software-development systems coordinate planners, coders, testers, and reviewers~\cite{hong2024metagpt,qian2023chatdev}, while information-seeking systems dispatch parallel researchers and integrate their results~\cite{chen2025mindsearch}.
Serving these applications therefore means executing concurrent multi-Agent workflows whose dependencies, model state, and resource demands unfold over time.

Agent workflows increasingly process confidential inputs and proprietary data, making private deployment a practical requirement in privacy-sensitive sectors~\cite{chrapek2025confidential,knoop2026private}.
Cost constraints keep these deployments on small, non-elastic GPU pools; fixed capacity and device heterogeneity make physical execution a first-order decision.
Agent roles invoke different models and request shapes, and the same activation can vary substantially in latency, memory demand, and feasibility across GPUs~\cite{wang2025lmmeter}.
A flexible review Agent can occupy the only high-capacity GPU required by a long-context synthesizer; model weights, KV caches, prefixes, and engine state can meanwhile remain useful to later activations.
Because model upgrades and incremental GPU procurement also leave sparse model--device measurements, the runtime must decide not only \emph{when} an Agent can run, but \emph{how} and \emph{where} to realize it.

Workflow progress exposes many physical choices before every downstream Agent becomes runnable.
Once its branch, role, model, and eligible devices are determined, the runtime can prefetch its deployment or retain and reclaim resident replicas while a predecessor still runs.
It can also fuse consecutive same-deployment activations, order independent units to avoid resource conflicts, and coordinate compatible model lifecycles across workflows.
These choices preserve the logical workflow but change loading, residency, peak memory, and critical-path time; poor choices instead evict useful models or occupy a scarce GPU.

Prior systems schedule or place agent calls~\cite{lin2024parrot,luo2025autellix,huang2026fate,wang2026maestro}, reduce model-loading overhead~\cite{fu2024serverlessllm}, anticipate KV reuse~\cite{pan2025kvflow,zheng2026pbkv}, or predict isolated executions~\cite{wang2025lmmeter}.
Existing planners couple several decisions but exclude admission-boundary transformations from joint lifecycle, ordering, placement, and cumulative-memory planning.
They therefore do not jointly evaluate a transformation's cumulative effects on ready work, scarce devices, memory, and resident state.

The runtime must expose each resolved cross-workflow window, forecast readiness and device-specific costs, construct legal physical alternatives, and build a feasible graph--schedule plan that adapts to workflow events and telemetry.

\textbf{Challenge 1: constructing legal physical executions.}
A logical agent graph fixes dependencies, roles, tool semantics, and externally visible effects, but leaves open how each activation is realized physically.
The runtime must construct a bounded set of useful same-deployment fusions and model-lifecycle realizations while preserving logical dependencies, request isolation, and completed results.

\textbf{Challenge 2: cross-device performance prediction.}
Physical-graph cost depends jointly on activation readiness, device-specific execution time, model-transition time, peak memory, and future deployment reuse.
Exhaustive profiling does not scale across models, devices, sequence lengths, and serving configurations, while underestimation can reverse a device ranking or make an infeasible timeline appear feasible.

\textbf{Challenge 3: feasible online graph--schedule construction.}
An alternative that is cheap in isolation can delay ready work, consume the only suitable GPU for a constrained agent, or violate capacity after model, request, engine, and transitional memory demands accumulate.
The runtime must choose graph structure, placement, resource order, and state lifetime together, commit only a small executable prefix, and revise the uncommitted suffix when arrivals, tool returns, or telemetry invalidate the plan.

\Projectname{} uses a resolved but not-yet-runnable workflow window to prepare physical execution without changing logical semantics.
At each scheduling event, \predictor{} forecasts activation--device costs and release times; \constructor{} enumerates legal fusion and model-lifecycle alternatives, including compatible actions shared across workflows; and \scheduler{} builds a feasible graph--schedule plan while protecting ready work.
\Projectname{} commits only through the next execution boundary and revises uncommitted physical decisions as workflow events and vLLM telemetry arrive, while preserving logical dependencies and results.

The main contributions of this work are\footnote{Code and data will be released upon acceptance.}:
\begin{itemize}
  \item \textbf{Graph construction.}
  We construct bounded physical alternatives for concurrent multi-agent workflows through same-deployment chain fusion and compatible model-lifecycle actions, and expose them for joint planning with resource ordering and placement.

  \item \textbf{Cross-device performance prediction.}
  \predictor{} estimates device-specific execution time, peak memory, and model-loading time and propagates these values through workflow dependencies to forecast activation readiness.

  \item \textbf{Adaptive scheduling.}
  \scheduler{} constructs a feasible physical graph and schedule under live pool state and cumulative memory constraints, protects ready work, and reranks affected uncommitted decisions using runtime feedback.
\end{itemize}

\section{Background and Motivation}
\label{sec:motivation}

\subsection{Agent Workflows and Serving State}

Multi-Agent workflows interleave model calls with inter-agent messages, tool results, and control-flow decisions, causing their execution graphs to emerge incrementally at runtime~\cite{wu2023autogen,luo2025autellix}.
We call each execution of a model node an \emph{activation}.
Each activation combines a role prompt, request input, and predecessor context; a repair round is therefore a new activation even when it reuses the same role and model.

As the graph unfolds, downstream activations often become visible before they can run.
An activation is \emph{ready} after all required predecessors complete and \emph{near-ready} when its branch, role, model, immutable input prefix, and eligible devices are known, every predecessor is complete or running, and at least one remains running.
A fan-in synthesizer can be near-ready while its inputs are produced, whereas a repair activation remains unknown until a test selects its branch.
Ready activations and immediate near-ready successors form a bounded window that exposes useful future structure without assuming unresolved branches.

Private multi-Agent deployments keep sensitive inputs within organizational infrastructure~\cite{chrapek2025confidential,knoop2026private}, where cost constraints limit spare capacity and elastic expansion.
Concurrent workflows therefore share a small, fixed GPU pool whose devices may differ in compute throughput, memory capacity, and resident state.
In this setting, early visibility matters because the same activation can incur different physical costs across devices.
The logical graph determines when a result may be consumed, while resident weights, KV cache, prefixes, engine state, and device capacity determine how the activation can be realized.
Retaining a deployment for a visible near-ready successor can avoid reloading; without one, the replica merely occupies scarce memory.
Workflow progress therefore links readiness to model residency and device placement, enabling planning beyond individual requests.

\begin{figure}[t!]
  \centering
  \includegraphics[width=\columnwidth]{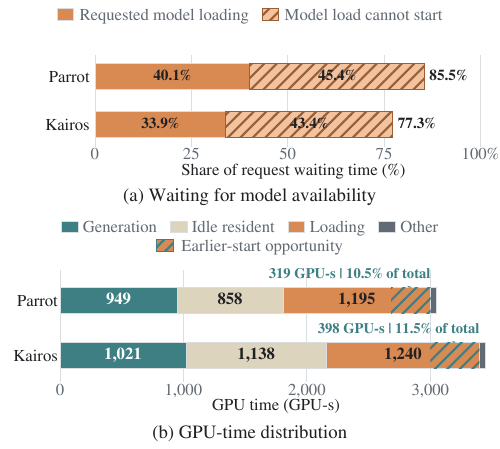}
  \caption{Model-lifecycle inefficiency under request-only scheduling.
  (a) Waiting for model availability.
  (b) GPU-time distribution; hatching marks loading that can start before request readiness.}
  \Description{Two panels compare Parrot and Kairos.
  Panel (a) shows that requested-model loading and inability to start model loading together account for 85.5 percent of Parrot waiting time and 77.3 percent of Kairos waiting time.
  Panel (b) reports absolute GPU-seconds: Parrot totals 3,052 GPU-seconds, with 949 generation, 858 idle resident, 1,195 loading, and 50 other; Kairos totals 3,449 GPU-seconds, with 1,021 generation, 1,138 idle resident, 1,240 loading, and 50 other.
  Hatching covers 319 and 398 GPU-seconds of loading, respectively, equal to 10.5 and 11.5 percent of total GPU time.}
  \label{fig:model-residency-motivation}
\end{figure}

\subsection{Limits of Request-Level Execution}
\label{sec:motivating-experiments}

Existing workflow schedulers expose dependencies or prioritize requests but still act on individual model calls once they become ready~\cite{lin2024parrot,tan2025ayo,Chen2025kairos}.
They can reorder ready work, but do not jointly manage the backend working set or preserve physical continuity across logical node boundaries.
We isolate these two limitations on a common serving substrate.

\paragraph{Workflow scheduling remains decoupled from model lifecycle management.}
Workflow information can change which ready request runs next, but not when the model required by a downstream node becomes available.
We evaluate Parrot and Kairos on a common serving substrate and measure the resulting model-availability delay and lifecycle overhead.

Figure~\ref{fig:model-residency-motivation}(a) decomposes model-availability delay into time spent waiting for an ongoing load and time during which the required load cannot start; together they account for 77.3--85.5\% of request waiting under both baselines.
Changing request priority therefore does not remove the dominant source of waiting.
Figure~\ref{fig:model-residency-motivation}(b) shows that model loading consumes more GPU time than token generation, while idle residency occupies another 28.1--33.0\% of total GPU time.
The hatched segments identify loading that workflow dependencies expose before request readiness, accounting for 10.5--11.5\% of total GPU time.
This interval is an opportunity bound rather than a measured speedup.
The limitation is therefore not a particular request priority, but the separation between workflow scheduling and model lifecycle management.
Addressing it requires treating model availability as part of workflow execution rather than as a reaction to each ready request.

\paragraph{Logical node boundaries create repeated physical admission.}
Dependency-aware scheduling still treats each model call as an independent scheduling unit.
When adjacent nodes require identical model and serving configurations, every successor releases, requeues, and reacquires the backend.
We measure this boundary for Parrot and Kairos while holding placement, admission, batching, and model-lifecycle mechanisms constant.
QMSum has two parallel three-node Qwen3-4B lanes followed by three Qwen3-8B aggregation nodes; six of nine calls have a same-model successor~\cite{zhong2021qmsum}.
A concurrent four-model MBPP repair workflow creates residency pressure on one A100-40GB and two V100-32GB GPUs~\cite{austin2021program}.
Three repetitions yield 30 QMSum sessions per baseline.

\begin{figure}[t!]
  \centering
  \includegraphics[width=\columnwidth]{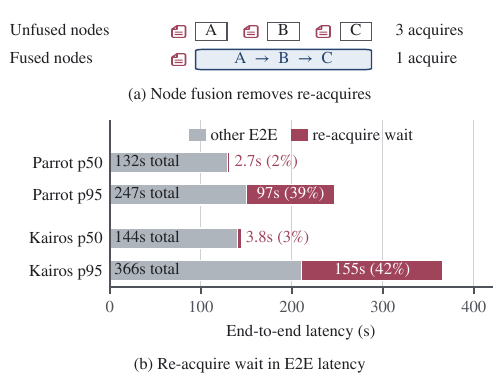}
  \caption{Repeated admission at same-model node boundaries.
  Highlighted segments are fixed-trace contributions of successor acquire waits, not measured fusion speedups.}
  \Description{The first panel contrasts three independent acquire operations with one acquire for a fused three-node sequence.
  The second panel shows stacked horizontal bars for Parrot and Kairos at p50 and p95.
  Re-acquire wait is two to three percent of median latency but thirty-nine to forty-two percent of p95 latency.}
  \label{fig:node-fusion-motivation}
\end{figure}

Figure~\ref{fig:node-fusion-motivation} shows a small median component---2.7--3.8\,s, or 2--3\%---but a large tail component: 97.1--155.1\,s, or 39--42\%, in the p95 sessions.
Trace-confirmed loading accounts for 65.9\,s and 130.7\,s of the two p95 segments.
Both baselines repeat admission at the same logical boundaries, identifying the physical boundary as the source of the delay.
Fusion changes the admission unit: the chain competes at its head and retains separate generations under one backend admission.
The fixed-trace components measure this opportunity; Section~\ref{sec:eval-node-fusion} reports the resulting net speedup under changed residency and interference.

\subsection{Opportunity for Physical-Graph Execution}
\label{sec:physical-opportunity}

The experiments expose two axes absent from request scheduling: state lifetime and the physical scheduling unit.
The realized logical window fixes dependencies and visible results, but not scheduling-unit boundaries, model-lifecycle actions, resource order, or placement.
These choices form a mutable physical execution graph.

\paragraph{Near-ready preparation.}
Near-readiness makes model weights, serving configuration, and an immutable prefix determinate before the remaining predecessors complete.
This interval exposes a preparation stage; \scheduler{} admits it only from residual capacity.
Its generation and any external side effect still wait for the original logical dependencies.
Consecutive nodes with identical model and serving configurations form one admitted unit while retaining separate prompts, generations, results, and commit points.
Across workflows, requests with matching model and serving configurations and authorized isolation share one inference backend and its model residency; private KV state, completion events, and semantic graphs remain separate.
Preparation, fusion, and cross-workflow sharing preserve workflow dependencies while changing physical realization.

\paragraph{Evaluate choices against cumulative heterogeneous state.}
A transformation that appears cheap in isolation becomes infeasible when its cumulative profile exceeds pool capacity.
Loading exceeds memory when old weights, retained KV, engine buffers, and incoming state coexist, even if the final set fits.
Placing a flexible activation on the only high-capacity GPU blocks a constrained successor; resource ordering preserves that device and reduces replacement and loading work.
vLLM further reacts to KV pressure through queueing, preemption, and recomputation~\cite{kwon2023vllm,vllm2026optimization}.
Physical alternatives therefore require device-specific, time-varying resource profiles that capture their cumulative effects.

The runtime must predict sparse model--device profiles, construct bounded legal alternatives across workflows, and jointly select graph structure, placement, resource order, and state lifetime under live pool state.
It must also limit commitment: tool returns, branch outcomes, completions, and serving telemetry invalidate future costs while the logical workflow remains unchanged.
\predictor{}, \constructor{}, and \scheduler{} address these three requirements as the prediction, construction, and adaptive-selection stages of \projectname{}.

\section{Design}
\label{sec:design}

\subsection{Overview}

\begin{figure}[t!]
  \centering
  \includegraphics[width=\columnwidth]{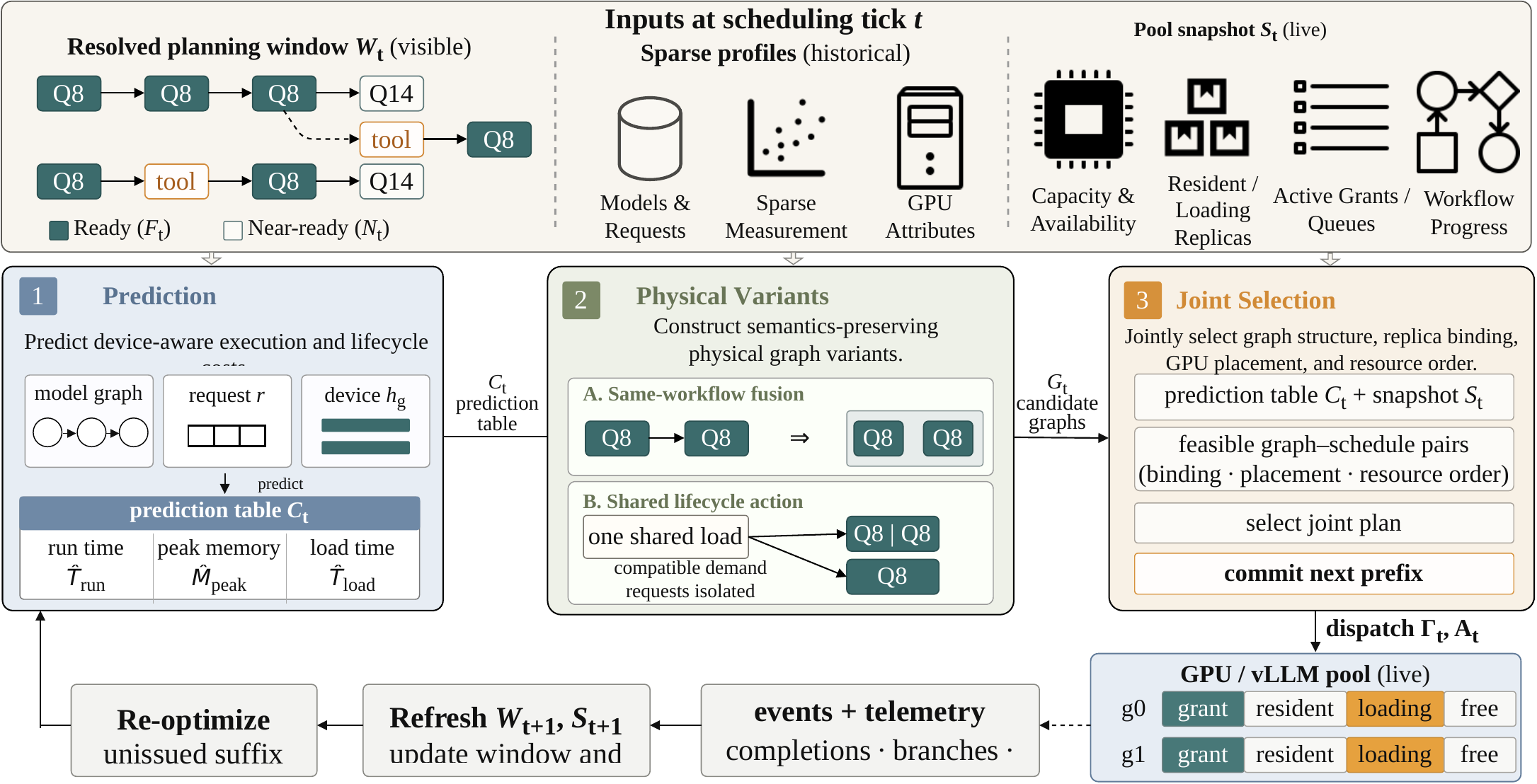}
  \caption{\Projectname{} combines semantics-preserving same-workflow fusion, shared lifecycle actions across isolated workflows, device-aware predictions, and live pool state to construct and incrementally commit a feasible graph--schedule plan.}
  \Description{Two concurrent Agent workflows feed Predictor, Constructor, and Scheduler. Predictor predicts execution time, peak memory, and loading time. Constructor fuses a semantics-preserving same-deployment chain within one workflow and shows a compatible lifecycle action shared by isolated requests from different workflows. Scheduler constructs a feasible graph--schedule plan, commits its next prefix to the GPU and vLLM pool, and replans from the updated pool snapshot.}
  \label{fig:architecture}
\end{figure}

Each multi-Agent workflow specifies what must happen and which results become visible, but leaves open how its model calls should share execution boundaries, replicas, or GPUs. Figure~\ref{fig:architecture} shows how \projectname{} jointly turns these logical workflows into an executable graph--schedule plan. \predictor{} estimates run time, peak memory, and loading time for each model--request configuration on every GPU type. Guided by these estimates, \constructor{} fuses compatible same-deployment chains and exposes reusable or shared lifecycle actions without changing workflow semantics. \scheduler{} incrementally constructs a feasible physical graph and schedule by ranking lifecycle, replica, placement, and resource-order decisions on a common timeline, then commits the next executable prefix. As calls complete and replica state changes, the same process revises the uncommitted suffix.

\subsection{Logical--Physical Execution Model}
\label{sec:execution-model}
\label{sec:workflow-definition}
\label{sec:execution-path}

At time $t$, the scheduler observes the resolved portion $L_t^w=\langle V_t^w,E_t^w\rangle$ of each active workflow $w\in\Omega_t$, where each activation is a distinct invocation. \constructor{} converts their union $L_t=\biguplus_{w\in\Omega_t}L_t^w$ into the fusion-normalized graph $G_t^0$. Let $F_t$ and $N_t$ denote its ready and near-ready units. A unit belongs to $F_t$ when all predecessors have completed; it belongs to $N_t$ when every predecessor has completed or is running and at least one remains running. Completion estimates for running predecessors come from the scheduler's committed timeline. These sets define the planning window $\mathcal W_t=G_t^0[F_t\cup N_t]$. A branch enters only after its control result resolves, so the window contains work on determined logical paths.

The window identifies what may execute next, but not how deployments become available or where units run. This distinction separates the logical frontier exposed by workflow progress from the physical choices made by the runtime. Let $U_t^0=V(\mathcal W_t)$. A physical realization is $G_t^p=\langle U_t^0\cup H_t^p,E_t^{L,p}\rangle$.
Here, $H_t^p$ contains device-symbolic loading, prefetching, and reclamation actions, while $E_t^{L,p}$ preserves logical dependencies and places each action before its consumers. The same logical window may therefore admit several realizations: reuse a resident replica, load a new one, or reclaim an idle replica before loading. The graph prescribes the required lifecycle actions; its schedule $\pi$ assigns devices, replicas, and reclamation victims, orders resource conflicts, and predicts start and completion times.

Snapshot $S_t$ records GPU capacities, resident and loading replicas, active grants and queues, and workflow progress. A deployment identity $d$ combines a resolved model and version with its serving configuration. Unit $u$ requires identity $d(u)$ and model $m(u)=m(d(u))$; units with the same identity may reuse a replica while retaining per-activation request state and outputs.

An execution grant admits a unit to a selected replica, and its lease prevents that replica from being reclaimed until the unit completes. The next grant completion or model-lifecycle transition forms an execution boundary, where $S_t$ is refreshed and the unissued suffix may be revised.

The invariants above define an admissible domain $\mathbb G_t$: every member is acyclic, preserves logical dependencies and visible values, respects deployment identity, and orders lifecycle actions before consumers. At each tick, \scheduler{} incrementally constructs a feasible $(G_t,\pi_t)$ and dispatches its next executable prefix; arrivals, completions, predecessor updates, and model transitions refresh the window and rerank the uncommitted work.

\subsection{\predictor{}: Device-Aware Prediction}
\label{sec:predictor}
\label{sec:lifecycle-profiles}

Rather than predicting workflow-level latency, \predictor{} maps model structure $\mathcal G_m$, request configuration $r$, and device attributes $\mathbf h_g$ to
\begin{equation}
  \label{eq:prediction-table}
  \left(
    \widehat T_{m,r,g}^{\mathrm{run}},
    \widehat M_{m,r,g}^{\mathrm{peak}},
    \widehat T_{d,g}^{\mathrm{load}}
  \right)
  =
  \operatorname{Predict}(\mathcal G_m,r,\mathbf h_g).
\end{equation}
Here, $d$ denotes the deployment identity for model $m$.
The mapping yields consistent execution-cost estimates for each activation across model configurations and eligible devices, while separating request-dependent execution costs from deployment-specific model-loading overhead.
Run time supports placement and completion-time comparison, peak memory supports admission, and loading time prices loading, prefetching, and reclamation.

\paragraph{Graph--device prediction.}
\predictor{} represents each model as an operator data-flow graph. Nodes describe operator type, arithmetic work, parameter size, and tensor shapes, while edges capture tensor dependencies and transferred data. Request features characterize the serving phase, batch size, and sequence lengths; device features characterize compute capability, memory capacity, and bandwidth. A graph encoder exchanges information along tensor edges and combines the pooled graph representation with the request and device context. Target-specific heads produce the three estimates in Equation~\ref{eq:prediction-table}, capturing request-dependent execution shapes and performance differences across heterogeneous GPUs.

\paragraph{Cached profiles and runtime feedback.}
Offline inference materializes run-time and memory profiles $\mathcal C_\kappa$ indexed by $\kappa=(d,g,b,\ell^{\mathrm{in}},\ell^{\mathrm{out}})$, together with loading profiles $\mathcal C^{\mathrm{load}}_{d,g}$. At runtime, the input lookup selects the smallest bucket covering the observed prompt. Timing uses the smallest output bucket covering the empirical 90th-percentile output length of recent completions of the same logical workflow node, falling back to its configured limit when no history exists. Admission always uses the bucket covering the configured output limit. Both paths therefore use cached lookups rather than executing the graph encoder on the scheduling path.

The memory reserved for admission is $M_\kappa^{\mathrm{adm}}(t)=\widehat M_\kappa^{\mathrm{peak}}+\epsilon_M+\rho_\kappa(t)$, where $\epsilon_M$ covers unmodeled variation and $\rho_\kappa(t)$ increases after an observed out-of-memory failure. Runtime history can therefore refine completion estimates without reducing the memory reserved for feasibility.

\paragraph{Forecasting.}
Execution estimates become workflow-level timing information by propagating them through logical dependencies. The predicted release of unit $u$ is
\begin{equation}
  \label{eq:predicted-release}
  \widehat r_u
  =
  \begin{cases}
    t, &u\in F_t,\\
    \displaystyle
    \max_{v\in\operatorname{pred}_{G_t^0}(u)}\widehat c_v,
      &u\in N_t.
  \end{cases}
\end{equation}
\constructor{} sums the predicted run times of activations fused into one scheduling unit. \scheduler{} then combines release time, loading, execution, replica queues, and resource precedence into candidate-specific completion times. Observed completions update downstream releases and ordering decisions, while live pool state and cumulative-memory checks determine final feasibility.

\subsection{\constructor{}: Physical-Graph Construction}
\label{sec:constructor}
\label{sec:execution-graph-optimizer}

Given the planning window and \predictor{}'s device estimates, \constructor{} forms physical candidates in two stages. It first contracts logical nodes that can share a backend grant within each workflow; after fusion, it derives the lifecycle actions required by the resulting units and coalesces compatible actions across workflows. For $L_t=(V_t,E_t)$, an edge $(u,v)$ belongs to a maximal fusible chain when
\begin{equation}
  \label{eq:fusion-condition}
  \begin{aligned}
  u,v\in V_{\mathrm A},\quad
  \operatorname{succ}_{L_t}(u)&=\{v\},\quad
  \operatorname{pred}_{L_t}(v)=\{u\},\\
  d(u)&=d(v),\quad \theta(u)\sim_{\mathrm{cfg}}\theta(v).
  \end{aligned}
\end{equation}
Here, $V_{\mathrm A}$ denotes Agent nodes, $d(\cdot)$ the deployment identity, and $\sim_{\mathrm{cfg}}$ configuration compatibility up to output-length limits. The one-to-one condition keeps branches and joins at chain boundaries. Contracting each maximal eligible chain yields $G_t^0=\operatorname{Fuse}(L_t)$. For $c=(v_1,\ldots,v_h)$, external dependencies attach to its endpoints, while $v_1\prec\cdots\prec v_h$ executes under one grant and replica lease. Each activation retains its request configuration and passes its output to the next, preserving the original data flow.

In QMSum, three consecutive aggregation nodes use the same Qwen3-8B deployment and become $c=(v_1,v_2,v_3)$. The fused unit requires one grant on one replica, which remains available across all three activations, while their prompts, generations, and logical outputs remain distinct.

Because an earlier activation's output may become part of a later activation's input context, admission covers the largest context exposed along the complete chain. For activation $v_i$, let $P_i$ be a bound on its fixed tokenized input, let $\mathcal A_i\subseteq\{1,\ldots,i-1\}$ index the earlier outputs it consumes, and let $O_i$ be its output limit; its input envelope is $\overline I_i=P_i+\sum_{j\in\mathcal A_i}O_j$. \constructor{} assigns
\begin{equation}
  \label{eq:fused-profile}
  \begin{aligned}
    I_c^{\mathrm{adm}}&=\max_{1\le i\le h}\overline I_i,
    \quad O_c^{\mathrm{adm}}=\max_{1\le i\le h}O_i,\\
    \widehat T^{\mathrm{run}}_{c,g}
      &=\sum_{i=1}^{h}\widehat T^{\mathrm{run}}_{m(c),r_i,g}.
  \end{aligned}
\end{equation}
Here, $r_i$ is the request configuration of $v_i$. The bounds $I_c^{\mathrm{adm}}$ and $O_c^{\mathrm{adm}}$ govern lease admission, while the summed duration predicts release. A length-$h$ chain uses one grant and removes $h-1$ intermediate scheduling boundaries.

After fusion, \constructor{} defines how each unit's deployment can become available. For $X\in\{F,N\}$, let $\alpha_F=\operatorname{Load}$ for ready work and $\alpha_N=\operatorname{Prefetch}$ for near-ready work. The lifecycle alternatives are
\begin{equation}
  \label{eq:lifecycle-alternatives}
  \mathcal H_t^{X}(u)=\{\langle\alpha_X(d(u))\rangle,
  \langle\operatorname{Reclaim},\alpha_X(d(u))\rangle\}.
\end{equation}
Lifecycle actions remain unplaced until scheduling. Reclamation selects a lease-free victim; resident binding reuses a replica, while loading creates or replaces one.

Let $h_{u,\rho}=1$ when unit $u$ selects lifecycle sequence $\rho$, with $\sum_{\rho}h_{u,\rho}\le 1$. A unit without a sequence reuses a resident replica or is deferred. Combining these per-unit choices yields a physical realization: compatible transitions may share one device-symbolic action node, while distinct nodes represent replication.

\begin{figure}[t!]
  \centering
  \includegraphics[width=\columnwidth]{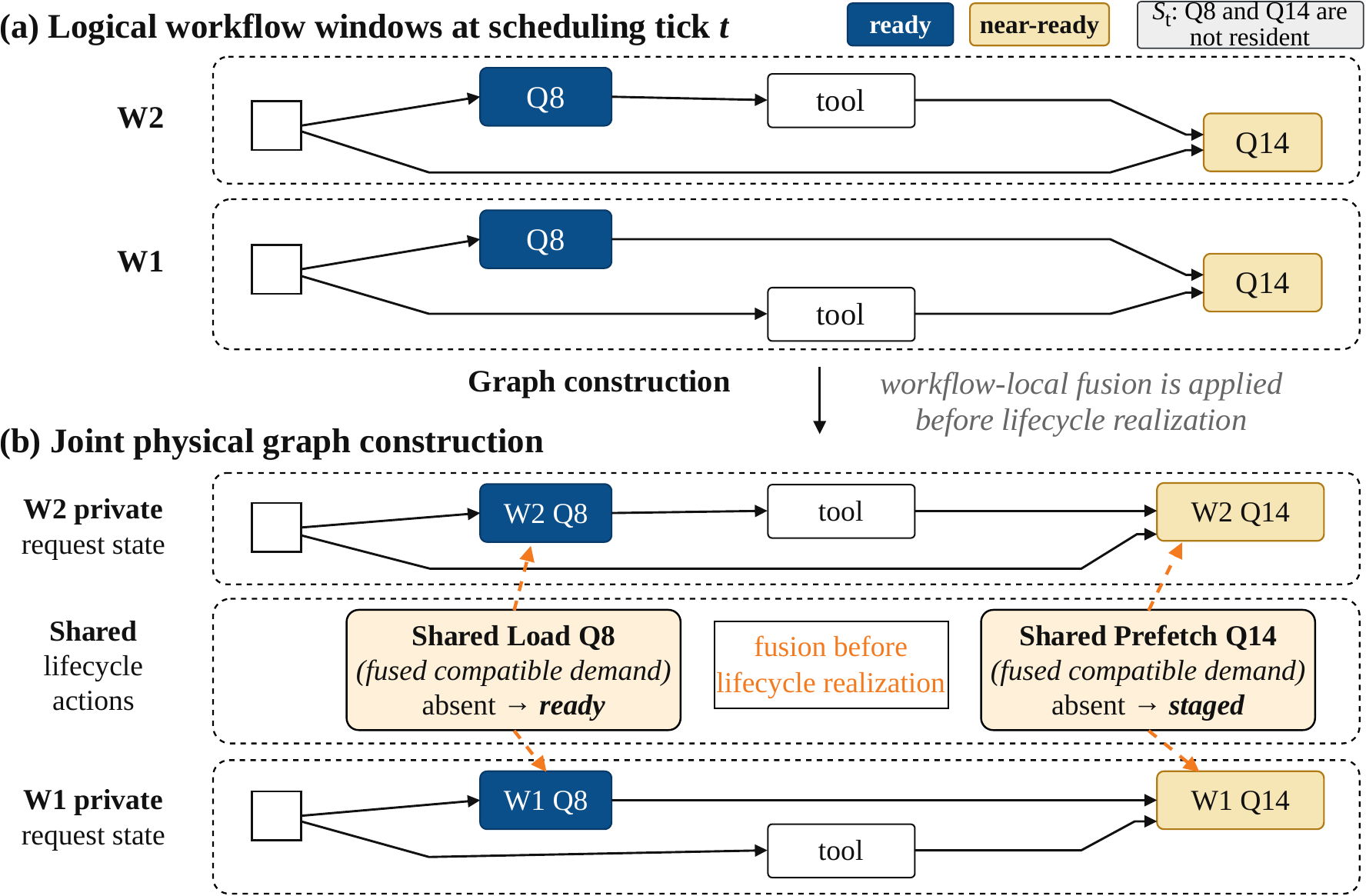}
  \caption{\constructor{} builds a joint physical graph at scheduling tick $t$. (a) W1 and W2 expose ready Q8 units and near-ready Q14 units while neither deployment is resident. (b) After workflow-local fusion normalization, compatible deployment demand across W1 and W2 shares one load that makes Q8 ready and one prefetch that stages Q14; request and tool state remain workflow-private.}
  \Description{Panel (a) shows the logical windows of W1 and W2 at scheduling tick t. Each contains a ready Q8 unit, a tool branch, and a near-ready Q14 unit; the snapshot records that Q8 and Q14 are not resident. Panel (b) shows separate W1 and W2 request and tool nodes consuming one shared Q8 load action and one shared Q14 prefetch action through dashed arrows.}
  \label{fig:joint-physical}
\end{figure}

Figure~\ref{fig:joint-physical} illustrates this construction order and separation. At tick $t$, W1 and W2 each expose a ready Q8 unit and a near-ready Q14 unit, while $S_t$ records both deployments as absent. \constructor{} first fusion-normalizes each workflow while preserving its logical dependencies. It then coalesces only compatible lifecycle demand across workflows: one load makes Q8 available to both ready units, and one prefetch stages Q14 for both near-ready units. The execution nodes and their request and tool state remain workflow-private.

Under snapshot $S_t$, let $\mathcal X_t^H$ collect the lifecycle selections and compatible action-sharing choices that satisfy the preceding constraint and the invariants in Section~\ref{sec:execution-model}. They induce the physical candidate space
\begin{equation}
  \label{eq:implicit-graph-space}
  \mathbb G_t
  =\left\{
    \operatorname{Realize}(\mathcal W_t,h):
    h\in\mathcal X_t^H
  \right\}.
\end{equation}
This implicit domain guides \scheduler{} in constructing one realization jointly with its schedule.

\subsection{\scheduler{}: Online Joint Graph--Schedule Construction}
\label{sec:scheduler}
\label{sec:lifecycle-orchestration}

\scheduler{} implements an event-driven scheduling heuristic. It chooses lifecycle actions, replica and device bindings, and conflict order together because the value of loading, reusing, or reclaiming a model depends on the resulting contention and completion time.

Consider a window in which two workflows expose ready units for deployment $d_1$, while a near-ready unit requires $d_2$. Reusing a resident $d_1$ replica avoids loading, but may serialize both ready units behind its queue. Loading another $d_1$ replica increases parallelism, yet occupies memory that could prefetch $d_2$; reclaiming $d_2$ may accelerate the ready work but incur a reload when the near-ready unit becomes eligible. Their effects become comparable once predicted releases, device-specific runtimes, replica queues, and memory transitions share one timeline. \scheduler{} ranks the resulting graph--schedule consequences on this timeline.

For operation $s\in V(G)$, let $\operatorname{pred}_{G}(s)$ and $\operatorname{rpred}_{\pi}(s)$ denote its graph and resource-order predecessors, and let $a_\pi(s)$ be its device. Its start follows both precedence relations,
\begin{equation}
  \label{eq:plan-timeline}
  \begin{aligned}
    \widehat s_s^{G,\pi}
    &=\max\!\left\{
      \widehat r_s,
      \max_{v\in\operatorname{pred}_{G}(s)}\widehat c_v^{G,\pi},
      \max_{v\in\operatorname{rpred}_{\pi}(s)}\widehat c_v^{G,\pi}
    \right\},\\
    \widehat c_s^{G,\pi}
    &=\widehat s_s^{G,\pi}
      +\widehat T_{s,a_\pi(s)}.
  \end{aligned}
\end{equation}
Here, $\widehat r_s$ follows Equation~\ref{eq:predicted-release} for units and is the current time for lifecycle actions. The duration uses \predictor{}'s run or load estimate, or the reclamation estimate recorded in $S_t$. This common timeline propagates model preparation and placement contention into the predicted completion of their consumers.

A candidate placement must first fit the deployment and request. Let $\mathcal D$ be the accelerators, with type $g(a)$ and memory capacity $C_a^M$; $I_u,O_u$ are unit $u$'s input and output bounds, and $\kappa(u,g,\nu)$ indexes its prediction at batch size $\nu$. Its feasible domain is
\begin{equation}
  \label{eq:placement-feasibility}
  \mathcal D_t(u,\nu)
  =
  \left\{
    a\in\mathcal D\ \middle|\
    \begin{aligned}
      I_u+O_u &\le L^{\max}_{d(u)},\\
      M^{\mathrm{adm}}_{\kappa(u,g(a),\nu)}(t)&\le C_a^M
    \end{aligned}
  \right\},
\end{equation}
where $L_{d(u)}^{\max}$ is the deployment's context limit and fused units use Equation~\ref{eq:fused-profile}. This domain is only the first feasibility cut. A fused unit can satisfy its context limit yet remain infeasible on a smaller GPU when its predicted admission memory exceeds capacity. Conversely, a device may enter the domain even when the deployment is not resident; the lifecycle choice and cumulative check below determine whether loading it fits the full timeline. Among the remaining devices, Scheduler compares completion times obtained from release, device availability, model preparation, and execution.

Each selected unit must either reuse a resident replica or follow a lifecycle action placed on the same GPU. For selected batch size $\nu_s$, write $\mathcal D_t(s)=\mathcal D_t(s,\nu_s)$; a shared action uses the intersection of its consumers' domains. Let binary $y_s$ select operation $s$, $x_{s,a}$ bind it to device $a$, and $\beta_{u,q}$ bind unit $u$ to an accepting replica $q\in Q_t(u)$ hosted by $a(q)$. Let $\lambda_u(G)$ indicate a loading or prefetch predecessor. These variables couple operation selection to device placement and replica supply. The graph--schedule consistency constraints are
\begin{equation}
  \label{eq:graph-plan-consistency}
  \begin{aligned}
    \sum_{a\in\mathcal D_t(s)}x_{s,a}&=y_s,
      &x_{h,a}&=x_{u,a}, &&(h,u)\in E^L(G),\\
    \sum_{q\in Q_t(u)}\beta_{u,q}+\lambda_u(G)&=y_u,
      &\beta_{u,q}&\le x_{u,a(q)}.
  \end{aligned}
\end{equation}
A reclamation action analogously binds one lease-free victim on its device.

Placement may introduce contention between logically independent operations. For each pair $(i,j,a)\in\mathcal K_t^\pi$ placed on the same nonconcurrent resource, the ranked action order induces one orientation:
\begin{equation}
  \label{eq:resource-order}
  \begin{aligned}
    z_{ij,a}+z_{ji,a}&=1,
      \qquad z_{ij,a},z_{ji,a}\in\{0,1\},\\
    z_{ij,a}=1&\Rightarrow \widehat s_j^{G,\pi}\ge\widehat c_i^{G,\pi},
      \quad
    z_{ji,a}=1\Rightarrow \widehat s_i^{G,\pi}\ge\widehat c_j^{G,\pi}.
  \end{aligned}
\end{equation}
The combined precedence relation remains acyclic.

Memory must be checked cumulatively because individually feasible operations can still overlap: a new replica may load while resident models and active requests remain live. At each predicted transition boundary $\tau\in\mathcal T_t^\pi$, let $R_{a,t}^{G,\pi}(\tau)$ be resident and transitional model memory, and $M_{s,a}^{\mathrm{inc}}$ the incremental request memory of active unit $s$. Memory feasibility requires
\begin{equation}
  \label{eq:cumulative-memory}
  R_{a,t}^{G,\pi}(\tau)
  +\sum_s x_{s,a}M_{s,a}^{\mathrm{inc}}
    \mathbf 1[\widehat s_s^{G,\pi}\le\tau<\widehat c_s^{G,\pi}]
  \le C_a^M,
  \quad \forall a,\tau,
\end{equation}
and concurrent units assigned to replica $q$ additionally respect its batch limit $B_q$. This transition-by-transition check rejects overlaps that exceed device capacity, turning a physical realization into an executable schedule that couples lifecycle choice, placement, resource order, and request execution on one predicted timeline.

Within this feasible region, \scheduler{} first exploits capacity that can serve ready work without another model load. A resident accepting replica provides immediate reuse; when loading is necessary, its value grows with the ready work it unlocks relative to predicted loading time, after accounting for priority and aging. Scale-out is considered only when a new replica completes the unit earlier than the current queues, subject to the replica cap. Pool-wide demand and queue state allow one residency decision to serve several workflows.

Reclamation considers idle, lease-free replicas, preferring redundant copies and then distant next use relative to reload cost; least-recently-used order handles missing estimates. A near-ready deployment becomes eligible for prefetch when its remaining release time approaches loading time plus dispatch slack, but only if prefetching does not delay ready work in the selected plan.

Planning remains tick-local: $\mathcal W_t$ contains only ready units and immediate near-ready successors. \scheduler{} processes ready units by priority and aging, first reusing accepting replicas and then choosing one feasible binding per remaining unit. Required loads, scale-out, and lease-safe reclamation are ranked by predicted benefit; near-ready prefetch is considered only when no ready action can advance. Resource conflicts inherit this action order, and each accepted decision is checked against Equations~\ref{eq:graph-plan-consistency}--\ref{eq:cumulative-memory}.

\scheduler{} ranks decisions first by ready-work coverage across priority classes, then by completion time, and finally by timely prefetching and lifecycle overhead. The map $k_t:F_t\rightarrow\{1,\ldots,K\}$ assigns each ready unit a priority and aging level. For the constructed plan, $A_k$ counts admitted level-$k$ units, $\widehat C_F$ is their latest completion, $P_N$ counts timely prefetches, and $H_{\mathrm{life}}$ combines ready-work loading delay with reclamation-induced reload cost. The lexicographic ranking key is
\begin{equation}
  \label{eq:joint-selection}
  \Phi_t(G,\pi)=
  \big(-A_1,\ldots,-A_K,\widehat C_F,-P_N,H_{\mathrm{life}}\big).
\end{equation}
Algorithm~\ref{alg:scheduler} applies this ordering while constructing a feasible plan and committing its next executable prefix.

\begin{algorithm}[t]
  \captionsetup{labelsep=colon}
  \caption{Online Joint Graph--Schedule Construction}
  \label{alg:scheduler}
  \small
  \renewcommand{\algorithmiccomment}[1]{\hfill$\triangleright$~\textit{#1}}
  \begin{algorithmic}[1]
    \REQUIRE Planning window $\mathcal W_t$; ready/near-ready sets $F_t,N_t$;
      snapshot $S_t$; prediction table $\mathcal C$
    \ENSURE Constructed plan $(G_t,\pi_t)$; grants $\Gamma_t$;
      lifecycle actions $\mathcal A_t$
    \STATE $R\gets\operatorname{RankReady}(F_t,S_t)$
      \COMMENT{Priority and aging.}
    \STATE $\Gamma_t,\mathcal L,\mathcal B\gets\varnothing,\varnothing,\varnothing$
    \FOR{$u\in R$}
      \STATE $b_u\gets\operatorname{BestFeasibleBinding}(u,S_t,\mathcal C)$
        \COMMENT{One binding.}
      \IF{$b_u$ reuses an accepting resident replica}
        \STATE $\Gamma_t\gets\Gamma_t\cup\{(u,b_u)\}$
      \ELSIF{$b_u\ne\bot$}
        \STATE $\mathcal L\gets\mathcal L\cup\{(u,b_u)\}$
      \ELSE
        \STATE $\mathcal B\gets\mathcal B\cup\{u\}$
      \ENDIF
    \ENDFOR
    \STATE $\mathcal A_t\gets
      \operatorname{PlanReady}(\mathcal L,\mathcal B,S_t,\mathcal C)$
      \COMMENT{Load, scale-out, or reclaim.}
    \IF{$\Gamma_t=\varnothing\land\mathcal A_t=\varnothing$}
      \STATE $\mathcal A_t\gets\operatorname{PlanPrefetch}(N_t,S_t,\mathcal C)$
    \ENDIF
    \STATE $(G_t,\pi_t)\gets
      \operatorname{Propagate}(\mathcal W_t,\Gamma_t,\mathcal A_t,S_t)$
      \COMMENT{Order conflicts; check feasibility.}
    \STATE $(\Gamma_t,\mathcal A_t)\gets\operatorname{Prefix}_t(G_t,\pi_t,S_t)$
    \STATE \textbf{return} $(G_t,\pi_t,\Gamma_t,\mathcal A_t)$
  \end{algorithmic}
\end{algorithm}

Algorithm~\ref{alg:scheduler} scans ready units in ranked order. For each unit, $\operatorname{BestFeasibleBinding}$ filters eligible devices and returns at most one binding. Resident replicas are reused first; remaining load, scale-out, and lease-safe reclamation actions are ranked by predicted benefit. Prefetch is considered only when no ready action can advance. $\operatorname{Propagate}$ serializes co-located operations and checks Equations~\ref{eq:graph-plan-consistency}--\ref{eq:cumulative-memory} before $\operatorname{Prefix}_t$ commits the next executable actions. Subsequent events refresh $\mathcal W_t$ and $S_t$ and rerank uncommitted work.

\section{Evaluation}
\label{sec:evaluation}

\subsection{Experimental Setup}
\label{sec:eval-setup}

\noindent\textbf{Testbed.}
The testbed comprises three heterogeneous nodes.
Each of two servers has two Intel Xeon Gold 6148 processors, with 20 cores per socket, and four NVIDIA V100 GPUs with 32\,GB of memory.
The remaining server has two Intel Xeon Silver 4216 processors, with 16 cores per socket, and four NVIDIA A100 GPUs with 40\,GB of memory.
All servers run CUDA 12 with the same software environment: vLLM 0.10.2, PyTorch 2.8.0, and Transformers 4.55.2.

\noindent\textbf{Models and Runtime.}
Five Qwen3 models (0.6B, 1.7B, 4B, 8B, and 14B) share one vLLM serving configuration~\cite{yang2025qwen3}.
Each resident replica occupies one GPU and serves compatible requests through continuous batching; replicas are loaded and reclaimed on demand.
All systems use the same inference engine, data type, and serving
configuration, so their differences arise from physical-graph
construction, lifecycle management, and scheduling.
Each arrival trace is generated once with seed 42 and replayed identically across all systems. Each system--trace configuration is executed five times, and we report the mean of the five run-level measurements.

\noindent\textbf{Workloads and Datasets.}
We construct three application-level workloads by pairing a public benchmark
with a representative Agent DAG. The datasets provide realistic task contents
and input-length distributions, while the DAGs determine inter-call
dependencies, parallelism, tool use, and model reuse.
GSM8K~\cite{cobbe2021gsm8k} contains diverse grade-school math problems requiring multi-step reasoning; its ensemble workflow runs five heterogeneous solvers in parallel, aggregates their answers by majority vote, and refines the result.
Sanitized MBPP~\cite{austin2021program} provides Python programming tasks with executable tests; its tool-augmented repair workflow performs code generation, testing, diagnosis, review, repair, and final validation while passing execution feedback across stages. The resulting workflow interleaves LLM and function nodes and contains both
sequential and multi-parent context dependencies, exercising handoffs among
four model deployments.
QMSum~\cite{zhong2021qmsum} contains query-focused summaries of long, multi-domain meeting transcripts; its hierarchical workflow uses two parallel three-stage rolling-summary chains followed by a three-stage aggregation chain.
Together, these workloads cover fork--join parallelism, tool-interleaved dependencies, long-context processing, fan-in aggregation, and repeated-model chains.

Each comparison uses identical workflow inputs, model assignments, arrival traces, and inference configurations.

\noindent\textbf{Baselines.}
The evaluation compares the default configuration of our system with two scheduling baselines and two predictor substitutions.
\begin{itemize}
  \item \textbf{Parrot}~\cite{lin2024parrot} introduces Semantic Variables that annotate request inputs and outputs, connect multiple LLM calls into a dataflow, and expose cross-request correlations to the serving system.

  \item \textbf{Kairos}~\cite{Chen2025kairos} combines a workflow-aware priority scheduler with a memory-aware dispatcher to reduce end-to-end latency when agents share loaded LLM instances.
\end{itemize}

\noindent\textbf{Formula predictor.}
This variant replaces \predictor{} with a static size-and-bandwidth estimator in the style of LLMFit~\cite{llmfit}; it derives runtime and memory estimates from model weights, KV-cache demand, and device bandwidth.

\noindent\textbf{GBDT predictor.}
This variant replaces \predictor{} with LightGBM~\cite{gbdt}, trained on scalar model, request, and device features including model size, layer count, batch size, sequence lengths, and memory bandwidth.

All five configurations use the same vLLM execution, placement, batching, and model-loading substrate.
The two predictor variants retain \constructor{} and \scheduler{} unchanged and therefore isolate prediction quality rather than define standalone schedulers.

\begin{figure}[t!]
  \centering
  \includegraphics[width=\columnwidth]{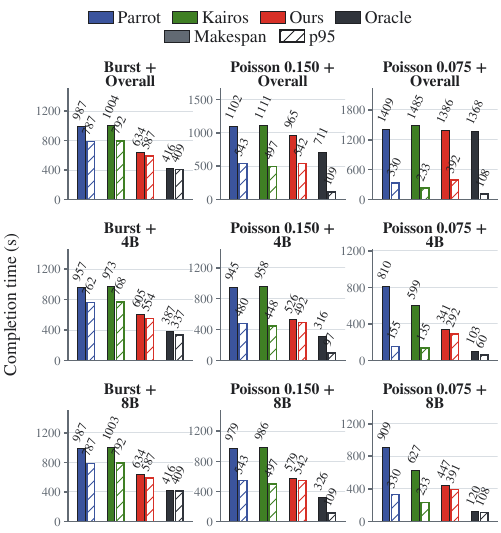}
  \caption{End-to-end completion time under burst and Poisson arrivals.}
  \Description{A three-by-three grid compares Parrot, Kairos, our system, and a trace-conditioned offline Oracle. Columns show burst and two Poisson arrival rates; rows show overall, Qwen3-4B, and Qwen3-8B completion. Each configuration has a solid makespan bar and a hatched p95 completion-latency bar; lower bars are better.}
  \label{fig:system-performance}
\end{figure}

\subsection{End-to-End Performance}
\label{sec:eval-end-to-end}

The workload contains 180 sessions, 60 per scenario.
Poisson arrivals use per-scenario rates of 0.150 and 0.075 sessions/s; the burst setting submits all sessions together.
Overall makespan measures the interval from the first arrival until the last session finishes; overall p95 is the 95th percentile of per-session completion latency.
Model-scoped completion metrics include DAG dependencies, queuing, model loading, and execution rather than only inference time.

The Oracle is a trace-conditioned offline scheduler with complete knowledge of task and model-loading durations.
It optimizes the displayed completion metrics subject to the same arrival trace, DAG dependencies, GPU capacity, model compatibility, and serving constraints as the measured systems.
We use it as a strong offline reference rather than a realizable online policy or a strict lower bound.

\Projectname{} finishes the workload in 634.2\,s, 35.8\% and 36.8\% earlier than Parrot and Kairos, respectively.
Its overall p95 is 587.0\,s, 25.4--25.9\% below the two baselines, and its 4B and 8B paths also finish first.

\Projectname{} also achieves the lowest overall makespan at both Poisson rates, finishing in 965.1 and 1386.4\,s at 0.150 and 0.075 sessions/s, respectively.
These results correspond to reductions of 12.4--13.1\% and 1.6--6.6\% over Parrot and Kairos.
Some Poisson p95 and model-scoped results favor another policy: \projectname{} optimizes global completion and can trade a local workflow tail for earlier workload drain.
Thus, \projectname{} drains the workload earlier in every arrival setting, although its global objective does not always minimize per-session or model-scoped tails under dispersed arrivals.

\subsection{GPU-Time Efficiency}
\label{sec:eval-gpu-efficiency}

We next use the same 180-session burst workload as Section~\ref{sec:eval-end-to-end} to report cluster-wide GPU time per completed session for each workflow scenario.
We attribute shared lifecycle time to scenarios by their share of each model's generation time.
We partition GPU time into generation, idle residency, model loading, and other activity.
Generation is time spent generating tokens, whereas idle residency is time during which loaded model weights occupy GPU memory without generating tokens.
Model loading is time spent preparing models, and other activity is the remaining GPU time.
Unlike a percentage-only breakdown, GPU time per completed session captures both how long GPU capacity is occupied and how that capacity is used.

\begin{figure}[t!]
  \centering
  \includegraphics[width=\columnwidth]{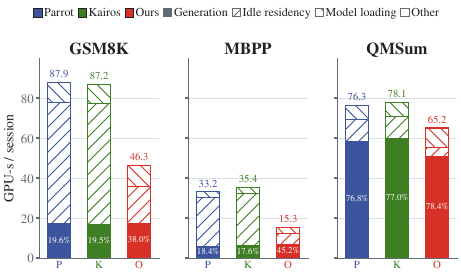}
  \caption{GPU time per completed session by workflow scenario under burst arrivals.}
  \Description{Three side-by-side panels compare Parrot, Kairos, and our system for GSM8K, MBPP, and QMSum on one shared GPU-second scale. Generation uses each system's solid color and contains its generation-share label, while white hatched segments show idle residency, model loading, and other activity. Our system has the shortest total bar and the highest generation share in every scenario.}
  \label{fig:gpu-time-efficiency}
\end{figure}

Figure~\ref{fig:gpu-time-efficiency} shows that \projectname{} uses less GPU time for all three workflow scenarios.
For GSM8K and MBPP, it uses 46.26 and 15.33 GPU-s per completed session, 46.9--47.4\% and 53.8--56.7\% less than the two baselines, respectively.
For QMSum, it uses 65.25 GPU-s, a reduction of 14.5--16.5\%.
Averaged across all 180 sessions, \projectname{} saves up to 24.63 GPU-s per completed session over the two baselines.

The reduction comes primarily from idle residency.
\Projectname{} lowers it from 60.32--60.68 to 18.27 GPU-s for GSM8K, from 24.43--26.39 to 5.25 for MBPP, and from 10.74--10.92 to 4.31 for QMSum.

\subsection{Prediction Accuracy}
\label{sec:eval-predictor}

\scheduler{} consumes cached resource profiles to compare the cost of running an Agent task on different devices and determine whether that task fits an available execution window.
Because a numerical error need not change either choice, the evaluation measures both numerical fidelity and downstream decision fidelity at the predictor--scheduler interface~\cite{power-aware-serve}.
2,798 measured Qwen3 model deployment configurations were used to form the test set.

Weighted absolute percentage error (WAPE) measures the numerical fidelity of loading-time, runtime, and peak-VRAM predictions by summing absolute errors over the test set and dividing by the sum of their measured values.
For downstream decision fidelity, \emph{ordering} measures whether predicted and measured runtimes agree on which of two configurations is faster~\cite{efficient-llm-scheduling-by-learning-to-rank}.
\emph{Window fit} measures whether the prediction and measurement agree that a configuration can finish within a given window.
Both metrics compare choices induced by cached predictions with those induced by measured profiles.
The window is the median measured runtime of the held-out configurations, 48.85\,s.
Together, they provide a closer proxy than point-estimate error alone for how cached predictions affect \scheduler{}'s decisions, while Section~\ref{sec:eval-end-to-end} measures the realized system-level impact.

\begin{table}[t!]
  \centering
  \scriptsize
  \setlength{\tabcolsep}{1.0pt}
  \caption{Prediction fidelity and decision accuracy of the three predictors across different prediction targets.}
  \label{tab:predictor-isobudget}
  \begin{tabular}{lrrrrr}
    \toprule
    & \multicolumn{3}{c}{WAPE (\%) $\downarrow$}
    & \multicolumn{2}{c}{Decision accuracy (\%) $\uparrow$} \\
    \cmidrule(lr){2-4}\cmidrule(lr){5-6}
    Method & Loading & Runtime & Peak VRAM & Ordering & Window fit \\
    \midrule
    \predictor{} & \textbf{8.05} & \textbf{7.25} & 14.08 & \textbf{97.94} & \textbf{97.82} \\
    Formula & 27.21 & 15.29 & \textbf{11.98} & 96.37 & 96.82 \\
    GBDT & 24.35 & 11.05 & 15.70 & 96.53 & 97.18 \\
    \bottomrule
  \end{tabular}
\end{table}

Table~\ref{tab:predictor-isobudget} shows that \predictor{} achieves the lowest loading-time and runtime WAPE, at 8.05\% and 7.25\%, respectively. This reduces loading-time WAPE by 70.4\% over the formula-based predictor and 66.9\% over GBDT; the corresponding runtime reductions are 52.6\% and 34.4\%.
Although all three predictors exceed 96\% decision accuracy, \predictor{} achieves the highest fidelity: it reproduces 97.94\% of measured runtime orderings and 97.82\% of measured window-fit outcomes, improvements of 1.41--1.57 and 0.64--1.00 percentage points over the baselines, respectively.
The formula-based predictor has the lowest peak-VRAM WAPE; \predictor{} is 2.10 percentage points higher but combines competitive VRAM fidelity with the best runtime and decision accuracy.

\subsection{Predictor and Component Ablations}
\label{sec:eval-component-ablation}
\label{sec:eval-node-fusion}

The analysis uses the same 180-session workload and three arrival processes as Figure~\ref{fig:system-performance} to compare the full system against two predictor substitutions and three component ablations.
The predictor substitutions use the formula-based or GBDT predictor while leaving the rest of \projectname{} unchanged.
The component ablations disable \constructor{} node fusion, disable near-ready prefetching, or make lifecycle decisions per workflow.
The per-workflow lifecycle condition continues to share loaded replicas across workflows, but each workflow separately estimates reuse, orders waiting requests, and evaluates loading benefit.
All three component ablations retain \predictor{} and the remaining mechanisms of our system.

\begin{figure}[t!]
  \centering
  \includegraphics[width=\columnwidth]{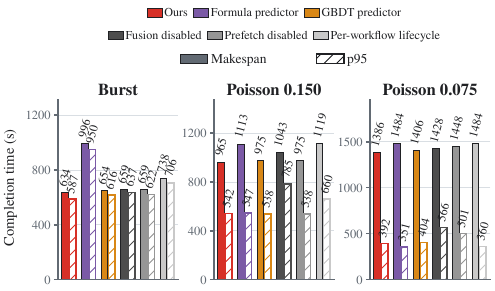}
  \caption{End-to-end performance of predictor substitutions and component ablations.}
  \Description{Three side-by-side panels compare our system, formula and GBDT predictor substitutions, fusion disabled, prefetch disabled, and per-workflow lifecycle conditions under burst, Poisson 0.150, and Poisson 0.075 arrivals. Each panel pairs solid makespan bars with hatched p95 completion-latency bars; lower bars are better.}
  \label{fig:component-ablation}
\end{figure}

Across the three arrival processes, the full system reduces makespan by 1.0--3.0\% over the GBDT substitution and 6.6--36.3\% over the formula-based substitution (Figure~\ref{fig:component-ablation}).
Among the component ablations, restricting lifecycle decisions to one workflow increases makespan by 7.1--16.4\%, while disabling node fusion raises p95 by 8.5--44.9\% and makespan by 3.0--8.1\%.
Disabling prefetching also increases makespan in every setting.
These results show that cross-workflow lifecycle coordination provides the largest makespan benefit, node fusion provides the strongest tail-latency benefit, and prefetching contributes consistently across arrival processes.

\section{Related Work}
\label{sec:related}

\noindent\textbf{Prediction.}
DNNPerf~\cite{gao2023dnnperf} uses a GNN to predict training time and GPU memory; PerfSeer~\cite{zhao2025perfseer} adds node, edge, and global features for multiple metrics.
nn-Meter~\cite{zhang2021nnmeter} sums kernel estimates on edge devices; Habitat~\cite{yu2021habitat} transfers measurements across GPUs; NeuSight~\cite{lee2025neusight} predicts unseen model--GPU pairs from hardware-bounded tiles.
HELP~\cite{lee2021help} adapts from a few target measurements; LitePred~\cite{feng2024litepred} transfers kernel predictors across similar platforms.
\predictor{} estimates run time, peak memory, and loading per model--request--device candidate for execution and lifecycle decisions.

\noindent\textbf{DAG execution and transformation.}
DryadLINQ~\cite{yu2008dryadlinq} compiles LINQ into distributed graphs; Tez~\cite{saha2015tez} supplies a DAG runtime.
Faastlane~\cite{kotni2021faastlane} co-locates serverless functions, while WiseFuse~\cite{mahgoub2022wisefuse} profiles DAG fusion and task bundling.
GMorph~\cite{yang2024gmorph} searches feature-sharing mutations across pretrained DNNs; Mercury~\cite{guan2025mercury} compiles remote-memory-aware multi-GPU operator schedules.
\constructor{} preserves Agent semantics while fusing deployment-identical chains and adding lifecycle vertices; \scheduler{} jointly selects bindings, placement, resource order, and cumulative memory.

\noindent\textbf{Serving-state management.}
vLLM~\cite{kwon2023vllm} uses PagedAttention for shareable KV caches; ServerlessLLM~\cite{fu2024serverlessllm} combines optimized checkpoints, tiered storage, and locality-aware scheduling to reduce startup latency.
ELORA~\cite{shi2026elora} jointly manages dependent LoRA--KV state; CrossPool~\cite{ye2026crosspool} disaggregates stable FFN weights and transient KV state.
KVFlow~\cite{pan2025kvflow} evicts and prefetches KV state; PBKV~\cite{zheng2026pbkv} predicts reuse; TokenCake~\cite{bian2025tokencake} protects critical-agent KV state and offloads it during tool calls.
Concurrent workflows additionally require state transitions to account for dependency-driven readiness, future model demand, and heterogeneous device availability across the pool.
\Projectname{} schedules model lifecycles against workflow reuse, contention, and cumulative pool occupancy.

\noindent\textbf{Agent serving and scheduling.}
Parrot~\cite{lin2024parrot} uses Semantic Variables to connect LLM-call producers and consumers, exposing inter-request dependencies to the scheduler.
Autellix~\cite{luo2025autellix} prioritizes calls by agent-program progress; Kairos~\cite{Chen2025kairos} combines workflow-aware priority with memory-aware dispatch.
Ayo~\cite{tan2025ayo} exposes primitive-level parallelism; Aragog~\cite{dai2025aragog} selects accuracy-equivalent routes under load.
Halo~\cite{shen2025halo} consolidates batched DAGs for computation, cache reuse, and placement;
SAGA~\cite{guo2026saga} schedules workflows using predicted cross-tool-call reuse; Maestro~\cite{wang2026maestro} coordinates model co-location, routing, and priority across clusters.
Most schedule ready calls or DAG units, leaving physical graph construction and future lifecycle demand outside a unified planning loop.
\scheduler{} jointly chooses fusion, lifecycle actions, binding, placement, and resource order under cumulative pool state and replans from feedback.
\section{Conclusion}
\label{sec:conclusion}

We present a prediction-guided runtime for concurrent Agent workflows on heterogeneous GPUs.
\Projectname{} separates logical semantics from state-dependent physical execution.
\predictor{} estimates activation--device costs and readiness across model
configurations and heterogeneous devices; \constructor{} enumerates legal fusion and model-lifecycle alternatives; and \scheduler{} selects and orders these alternatives under live pool state, revising uncommitted decisions from feedback.
Across a workload spanning three workflow scenarios, \projectname{} reduces burst makespan and overall p95 completion latency by up to 36.8\% and 25.9\%, respectively, over Parrot and Kairos, while saving up to 24.63 GPU-s per completed session.
These results show the value of joint physical-graph and lifecycle orchestration.

\begin{acks}
We use generative AI tools, including Writefull and ChatGPT, to support the proofreading and language refinement. These tools help us identify grammatical errors, improve clarity and readability.
\end{acks}

\bibliographystyle{styles/acmart-primary/ACM-Reference-Format}
\bibliography{references}

\end{document}